\documentclass[a4paper,11pt]{article}
\usepackage[left=1in,right=1in,top=1in,bottom=1in]{geometry}

\usepackage{authblk} % Inclusion of authors and affiliations
\usepackage{setspace} % Spacing of document
\usepackage{amsmath} % Used to align equations
\usepackage{graphicx} % Used to bring in figures

\usepackage[style=nature,url=false,isbn=false,doi=false]{biblatex}
\newcommand{\comment}[1]{} % Commannd to comment out any source in curly brackets

\begin{document}
\title{Cupolets in a Chaotic Neuron Model}
\author[1,3]{John E. Parker}
\author[2]{Kevin M. Short}
\affil[1]{Department of Mathematics, University of Pittsburgh, Pittsburgh, Pennsylvania 15260, USA}
\affil[2]{Integrated Applied Mathematics Program, Department of Mathematics and Statistics, University of New Hampshire, Durham, New Hampshire 03824, USA}
\affil[3]{Corresponding Author, jep220@pitt.edu}
\maketitle

\begin{abstract}
  This paper reports the first finding of cupolets in a chaotic Hindmarsh-Rose neural model. Cupolets (\emph{c}haotic, \emph{u}nstable, \emph{p}eriodic, \emph{o}rbit-\emph{lets}) are unstable periodic orbits that have been stabilized through a particular control scheme applying a binary control sequence. We demonstrate different neural dynamics (periodic or chaotic) of the Hindmarsh-Rose model through a bifurcation diagram where the external input current, $I$, is the bifurcation parameter. We select a region in the chaotic parameter space and provide the results of numerical simulations. In this chosen parameter space, a control scheme is applied when the trajectory intersects with either of the two control planes. The size of the control is determined by a bit in a binary control sequence. The control is either a small microcontrol (0) or a large macrocontrol (1) that adjusts the future dynamics of the trajectory . We report the discovery of many cupolets with corresponding control sequences and comment on the differences with previously reported cupolets in the double scroll system. We provide some examples of the generated cupolets and conclude by discussing potential implications for biological neurons.
\end{abstract}

\newpage

% INTRODUCTION
% Contains Introduction section of HR Single Cupolet Paper
\section{Introduction}\label{sec:introduction}
Throughout recent decades, many approaches have been taken to try to understand biological neurons. In this paper, we provide insight into neurons through examining the Hindmarsh-Rose (HR) mathematical model of neural firing. The HR neuron is represented by three coupled nonlinear ordinary differential equations, stated in Eq~\ref{eq:hr}. This model is particularly interesting for mathematical investigations since it is 3-dimensional and can exhibit several different modes of firing including chaotic behavior. \textcite{Izhikevich2004} provides a brief overview of the types of behavior the HR model can exhibit, as well as the possible behaviors of common neural models. The development of mathematical neural models that led to the the HR model is briefly outlined in the following paragraphs.

The seminal work of Hodgkin and Huxley in 1952 provided foundational connections between biology and the mathematics of an individual neuron \cite{Hodgkin1952}. The HH model is a 4-dimensional dynamical system, reflects the biophysics of a neuron, and is capable of simulating observed activity of a neuron. Numerical solutions of the HH model show the build up of the membrane potential of a neuron until a certain voltage threshold is reached. Then, the model neuron fires an action potential that can be seen as a spike in the membrane potential before entering a refractory period where the membrane voltage begins to build up again.

Many alternative models have since been proposed that reflect the complex dynamics of the HH model. One of the first was the FitzHugh-Nagumo (FHN) model \cite{FitzHugh1961,Nagumo1962} and was the focus of \textcite{Parker2020}. In \textcite{Parker2020}, the authors demonstrated how two chaotic FHN neurons, interacting bidirectionally, can transition to periodic firing through sigmoidal synaptic learning.

The FHN model is a simplification of the HH model from 4-dimensions to 2-dimensions, that is capable of exhibiting a single spike or periodic spiking firing modes. Even though the FHN model exhibits similar firing dynamics as the HH model, the parameters in the FHN model are not directly related to the biological or chemical properties of a neuron. Instead, one variable reflects the neuron's membrane potential and the other variable reflects the recovery dynamics of the neuron. Another 2-dimensional neural system, called the Morris-Lecar (ML) model \cite{Morris1981}, is also capable of similar dynamics as the FHN model, but the parameters of the ML model are more closely connected to the biological and chemical properties of the neuron. In 1985, \textcite{Chay1985} presented a 3-dimensional neural model as a reduced HH system similar to the Morris-Lecar model where the parameters reflect biophysical properties. The dynamics of this model are further explored in \textcite{Fan1994}. An extension of the FHN model was introduced by Hindmarsh and Rose as a 3-dimensional autonomous dynamical system, now known as the Hindmarsh-Rose (HR) model \cite{Hindmarsh1984}. Similar to the FHN model, the HR model parameters are not biochemically based, yet the overall dynamics are consistent with observed neural firing patterns. One key benefit of the added dimension of the HR model is that it allows for the potential for chaotic dynamics to occur, which will be the focus of the work reported here. For a reference on many commonly used mathematical neural models and the properties of each model, see \textcite{Izhikevich2004}.

Chaotic dynamics occur when the behavior of a dynamical system is aperiodic, remains bounded for all time, and is sensitive to initial conditions. Often this is characterized by a positive exponential rate of divergence between two close trajectories, commonly measured by the maximal Lyapunov exponent (MLE). The HH model is capable of exhibiting chaotic dynamics and an example of chaotic dynamics within the original parameters is discussed in \textcite{Guckenheimer2002}. \textcite{Rabinovich1999} suggest that chaotic behavior is important for connected networks of neurons as a means to produce desired rhythms, citing lobster central pattern generators (CPG) as an example. The authors then show how a network of HR neurons can model a CPG even though each individual neuron would have chaotic dynamics without the coupling. \textcite{Faure2001} and \textcite{Korn2003} discuss the role of chaos within the brain. \textcite{Erichsen2006} discuss the periodic and chaotic dynamics that occur when two HR neurons are coupled. \textcite{Shuai1999} explore phase synchronization between coupled HR neurons. Recently, \textcite{DoungmoGoufo2019} reported on the influence of an external current on the Hindmarsh-Rose system, focusing on numerical investigations into chaotic poles of attraction.

In this paper, we further examine the role of chaos in the HR neural model. Specifically, we examine how a single chaotic neuron can produce many periodic states dependent on received inputs. We discuss a control scheme applied to a chaotic HR neuron that causes the system to stabilize onto periodic orbits. These stabilized periodic orbits are highly accurate approximations of unstable periodic orbits. The unstable periodic orbits of a chaotic system create a skeleton of the chaotic attractor, and locally provide a good proxy for the evolution of chaotic trajectories as they evolve around the attractor~\cite{Cvitanovic1991}.

The control scheme that will be used here is adapted from \textcite{Hayes1993,Hayes1994} and is described in detail in \textcite{Parker1999, Zarringhalam2006, Zarringhalam2008, Morena2014b, Morena2014a, Morena2014, Short2019, Morena2020}. The control scheme is a discrete approach that applies small kicks to the chaotic system along specified control planes. These kicks are idealizations of the impulses that would be received from a signal coming up a chain of neurons. More specifically, the approach uses a repeated binary control sequence (e.g. 101100101100...) to apply relatively small perturbations or controls along a control plane. The type of control is designated by the selected bit in the control sequence, where a 0 bit would correspond to a tiny microcontrol and a 1 bit would be a macrocontrol that is the proxy for receiving an external pulse. This adapted control scheme has been applied to other dynamical systems, mainly the double scroll oscillator \cite{Matsumoto1985}. When certain control sequences are applied to a chaotic system via this control scheme, the chaotic behavior stabilizes onto persistent, periodic orbits called cupolets (\emph{c}haotic, \emph{u}nstable, \emph{p}eriodic, \emph{o}rbit-\emph{lets}). These cupolets were originally described in \textcite{Parker1999} and  discussed with applications in more detail in \textcite{Short2005, Short2005a, Zarringhalam2006, Zarringhalam2008, Morena2014, Morena2014a, Morena2014b, Short2019, Morena2020}. In this paper, we will demonstrate how this control scheme stabilizes chaotic HR behavior onto periodic, persistent cupolets.

The outline of this paper is as follows. In \S\ref{sec:model} we discuss the chaotic HR neural model and provide a visualization of the chaotic regime of interest. Then, in \S\ref{sec:control_scheme} we further explain the control sequence used to produce cupolets and illustrate properties of the system that result from applying controls. We conclude in \S\ref{sec:cupolets} by providing examples of some of the cupolets that were discovered and discuss implications of this work in \S\ref{sec:hr_discussion}.

% MODEL
% Contains Model section of HR Single Cupolet Paper
\section{Model}\label{sec:model}
\begin{figure}
  \centering
  \includegraphics[width=\textwidth,height=0.8\textheight,keepaspectratio]{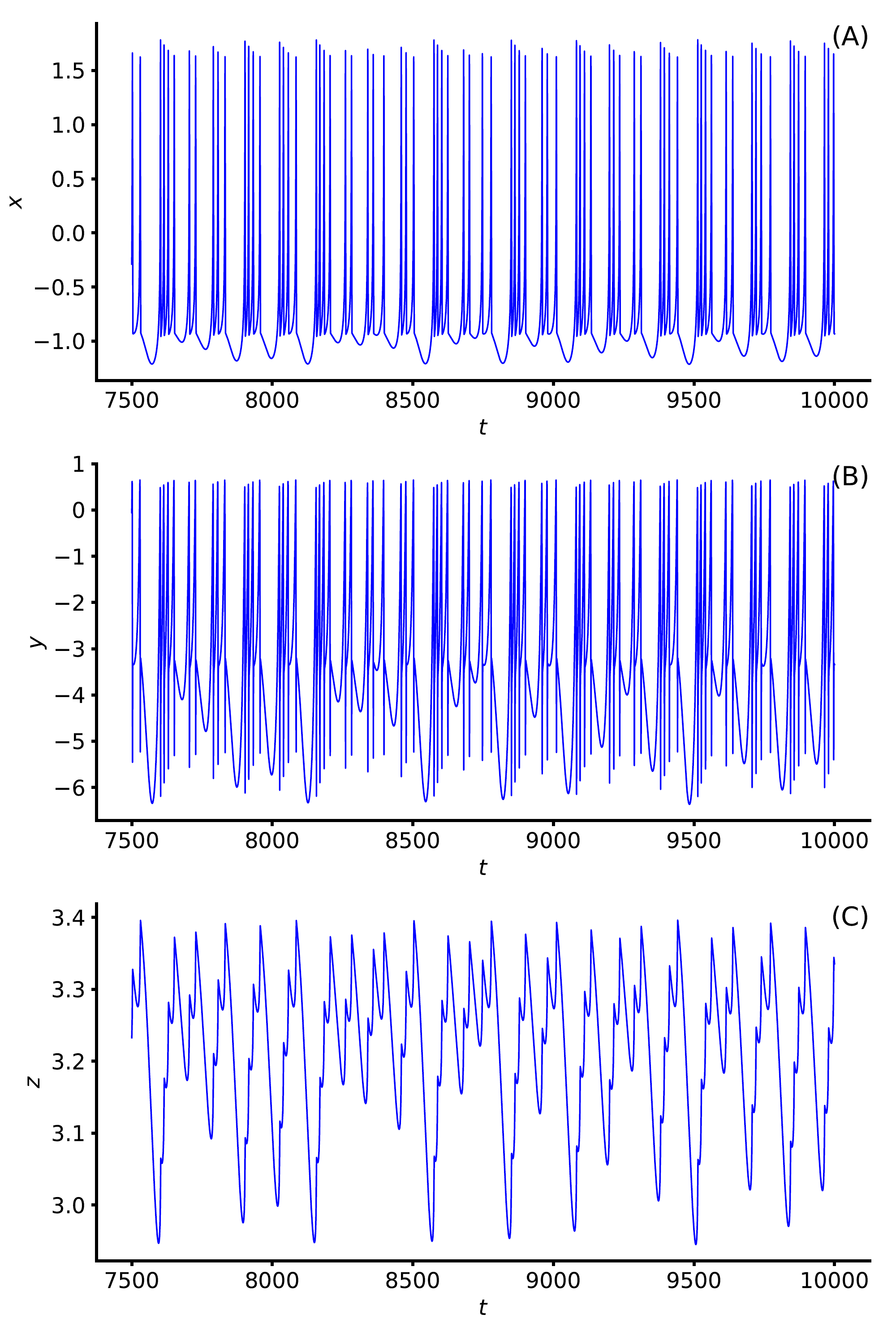}
  \caption{\label{fig:hr_chaotic_time_series}Numerical integration of Eq \ref{eq:hr} with the original parameters, $a=1$, $b=3$, $c=1$, $d=5$, $s=4$, $x_r = -8/5$, $r = 0.006$, and $I=3.25$. Numerical integration was performed using explicit the Runge-Kutta 4th order method with a fixed timestep of $dt=1/128$ for 1280000 iterations. The final 320000 iterations (25\% of numerical integration) are shown. (A) $x$ time series. (B) $y$ time series. (C) $z$ time series.}
\end{figure}

The neural system of interest is the Hindmarsh-Rose model and it has been chosen since it can exhibit chaotic behavior. Eq \ref{eq:hr} provides the original HR model as given in \textcite{Hindmarsh1984}. The original parameter values used are $a=1$, $b=3$, $c=1$, $d=5$, $s=4$, $x_r = -8/5$, with $r$ and $I$ varying. The parameter $I$ represents a direct current input to the dimensionless membrane potential, $x$. The neural membrane potential continuously absorbs input until a threshold is reached, thus firing an action potential and entering a refractory period. The dimensionless $y$ and $z$ variables represent the recovery dynamics governing the refractory period. Figure \ref{fig:hr_chaotic_time_series}A-C provide an example of the time series for the $x$, $y$, and $z$ variables, respectively.

\begin{equation}\label{eq:hr}
  \begin{aligned}
    \dot{x} &= y-ax^3+bx^2+I-z \\
    \dot{y} &= c-dx^2-y\\
    \dot{z} &= r(s(x-x_1)-z)
  \end{aligned}
\end{equation}

The $z$ recovery variable has a slower characteristic timescale than the $y$ recovery variable, and is controlled by the $r$ parameter. \textcite{Hindmarsh1984} discusses the different dynamics Eq \ref{eq:hr} can exhibit, including periodic firing, bursts, and random firing. Typically, Eq \ref{eq:hr} exhibits bursting behavior often seen in biological neural observations. That is, the $x$ variable has multiple consecutive spikes (action potentials) before entering the refractory period that is governed by the recovery dynamics ($y$ and $z$ variables). This bursting behavior cannot occur in the previously discussed FitzHugh-Nagumo neural model. \textcite{Hindmarsh1984} report random bursting with $r=0.005$ and $I = 3.25$. In this paper, we focus on a similar parameter set that produces chaotic dynamics where $r=0.006$ and $I=3.25$.

\begin{figure}
  \includegraphics[width=\textwidth,height=\textheight,keepaspectratio]{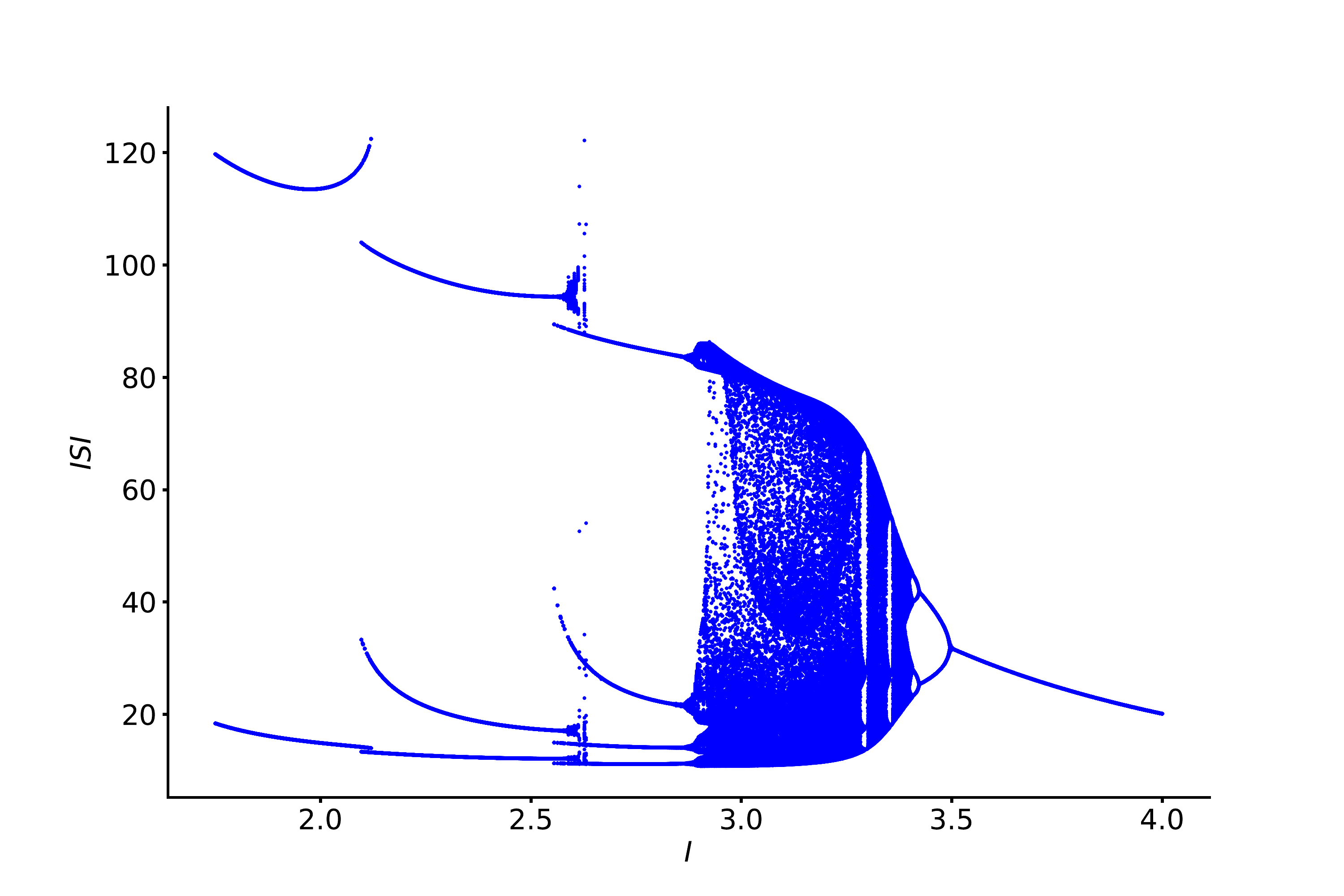}
  \caption{\label{fig:bifur}Bifurcation diagram for the HR model in Equation \ref{eq:hr}. The bifurcation parameter, $I$, is on the horizontal axis and varies from $I_{min}=1.750$ to $I_{max}=4.000$ , with $\Delta I = 0.001$. The vertical axis shows the interspike interval (ISI) of the $x$ variable in Eq \ref{eq:hr}.}
\end{figure}

This parameter set was selected after first conducting a bifurcation analysis. The bifurcation analysis allowed for examination of the different types of long-term spiking behavior (periodic or chaotic) with different $I$ values. In Figure \ref{fig:bifur}, a bifurcation diagram depicts the different long term behaviors for Eq \ref{eq:hr} as $I$ is varied. The bifurcation parameter, $I$, is on the horizontal axis ranging from $I_{min}=1.750$ to $I_{max}=4.000$ with $\Delta I = 10^{-3}$. The interspike interval (ISI) is on the vertical axis and measures the time between two consecutive spikes. At a given $I$ value on the horizontal axis, say $I=I_0$, a thick band of ISI values would indicate that Eq \ref{eq:hr} evolves chaotically (e.g. $I_0 = 3.25$), whereas if there are only a few ISI values plotted above a given $I_0$ value, then that indicates regular, periodic behavior. For example, in Figure \ref{fig:bifur} when $I < 2.5$ and $I > 3.5$ the HR model produces periodic behavior. \textcite{Innocenti2007} provides a more detailed analysis of the possible dynamics exhibited by varying $I$ and further explores chaotic dynamics of the HR neuron in \textcite{Innocenti2009}. \textcite{Storace2008a} analyze the bifurcations that occur when varying the two parameters $I$ and $b$.

\begin{figure}
  \includegraphics[width=\textwidth,height=\textheight,keepaspectratio]{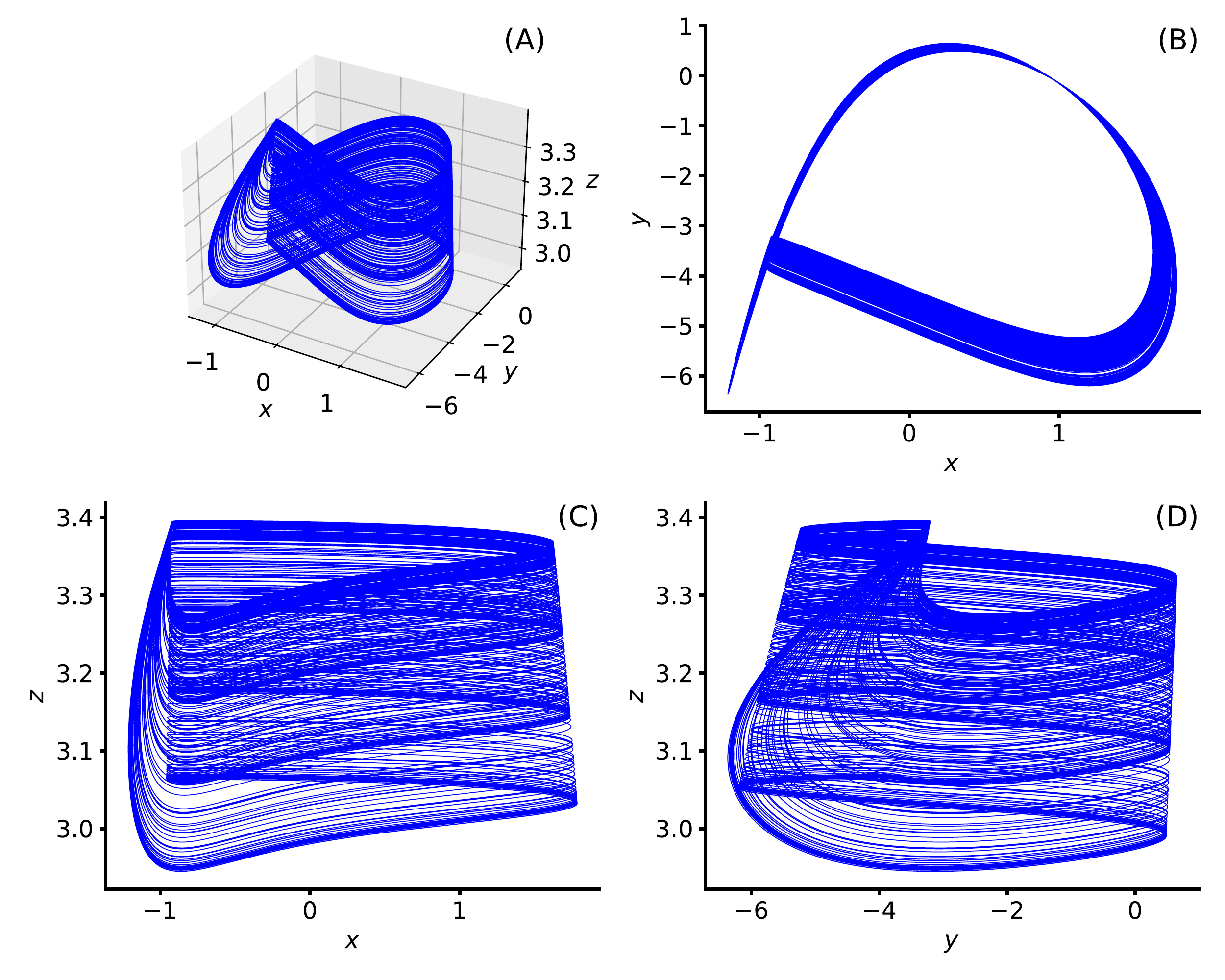}
  \caption{\label{fig:hr_chaotic_3d}Numerical integration of Eq \ref{eq:hr} with the parameters, $a=1$, $b=3$, $c=1$, $d=5$, $s=4$, and $x_r = -8/5$ from \textcite{Hindmarsh1984}. We choose $r = 0.006$ and $I=3.25$ with the parameters such that the model is in a chaotic region. Numerical integration was performed using the explicit Runge-Kutta 4th order method with a fixed timestep of $dt=1/128$ for 1280000 iterations. The final 960000 iterations (75\% of numerical integration) are displayed so that any transient behavior has passed. (A) 3-Dimensional phase space plot of $x$, $y$, and $z$ dynamics. (B) Projection of (A) into the $x$-$y$ plane. (C) Projection of (A) into the $x$-$z$ plane. (D) Projection of (A) into the $y$-$z$ plane.}
\end{figure}

From Figure \ref{fig:bifur}, when $I=3.25$ the HR model exhibits chaotic behavior. Figure \ref{fig:hr_chaotic_3d} shows the phase space resulting from numerical integration of Eq \ref{eq:hr} with these parameters while Figure \ref{fig:hr_chaotic_time_series} provides the corresponding chaotic time series. Figure \ref{fig:hr_chaotic_time_series}A shows the $x$ time series where the spikes per burst do not have any pattern and each burst consists either of 2, 3, or 4 spikes. Figure \ref{fig:hr_chaotic_time_series}B shows the $y$ time series, and Figure \ref{fig:hr_chaotic_time_series}C shows the $z$ time series. Neither Figure \ref{fig:hr_chaotic_3d} nor Figure \ref{fig:hr_chaotic_time_series} contain any periodic behavior.

All simulations in this paper were performed numerically using an explicit Runge-Kutta 4th order method with $dt = 1/128$. In Figure \ref{fig:hr_chaotic_3d} and Figure \ref{fig:hr_chaotic_time_series} the total simulation time of $t_{final}=10000$ gives 1280000 iterations, which provides enough time for the system to settle onto the attractor after any initial transient.

% CONTROL SCHEME
% Contains Control Scheme section of HR Single Cupolet Paper
\section{Control Scheme}\label{sec:control_scheme}
In this section we describe the control scheme for the HR neural system. The control scheme is adapted from the control scheme used by \textcite{Hayes1993,Hayes1994} on the double scroll oscillator and was used to generate cupolets in the references cited in \S\ref{sec:introduction}. This adapted control scheme uses two control planes that exist on two different sections of the attractor. Each control plane is placed such that the plane represents a certain stage of neural firing. Controls are applied only when the trajectory intersects with a control plane. The applied controls are proxies for the impulses a neuron would receive from nearby neurons. The control planes are lower dimensional (two-dimensional) Poincar\'e sections that are approximately orthogonal to the flow of the trajectory.

\begin{figure}
  \includegraphics[width=\textwidth,height=\textheight,keepaspectratio]{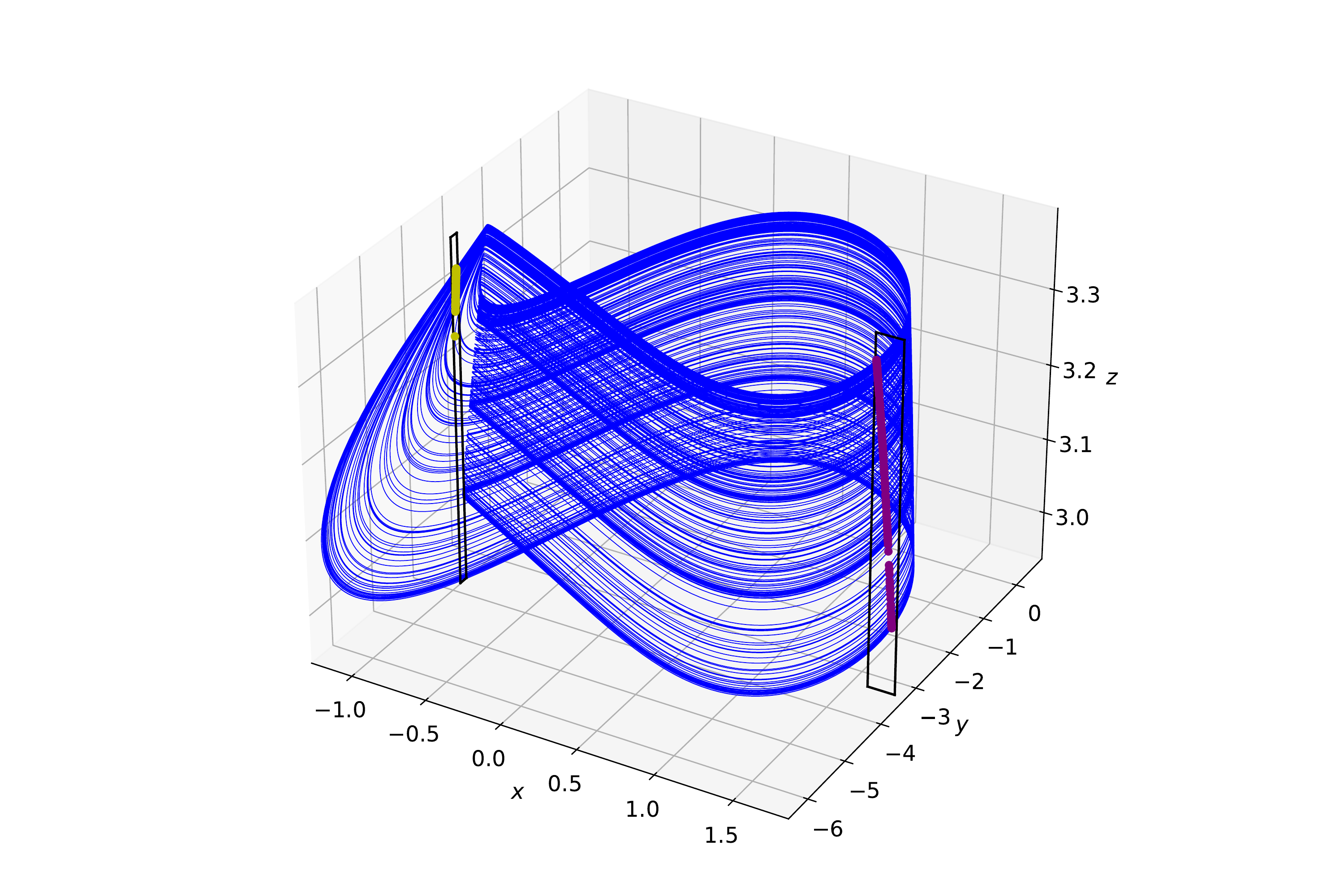}
  \caption{\label{fig:control_planes} Control planes of Eq \ref{eq:hr} after numerical integration. Each control plane is outlined as a black rectangle. The left control plane is PS0 and is defined as the plane corresponding to $x = -0.9832605683131186$. The yellow points inside PS0 are the points found exactly on the plane using Henon's trick \cite{Henon1982}. The PS0 intersections are in the upper section of the black rectangle outlining the control plane. The right control plane is PS1 and is defined as the plane corresponding to $y=-3.3657609537434663$. The purple points inside PS1 are the points found exactly on the plane using Henon's trick.}
\end{figure}

The control planes are chosen such that each control plane is placed in a position that corresponds to a key aspect of the bursting behavior of the system. One control plane, called Poincar\'e section 0 (PS0), is defined as the plane corresponding to the average local minima of the $x$ time series of each spike. For the numerical integration in Figure \ref{fig:hr_chaotic_3d} this results in $x = -0.9832605683131186$. This plane encompasses the refractory dynamics of the neural system that occurs after firing a burst of spikes. The second control plane, called Poincar\'e section 1 (PS1), is defined as the plane corresponding to the average $y$ value of the average $x$ local maxima. For the numerical integration in Figure \ref{fig:hr_chaotic_3d} this results in $y=-3.3657609537434663$. After numerical integration of Eq \ref{eq:hr} to determine the control planes, Henon's trick \cite{Henon1982} is used to find the exact points of intersection of the trajectories with each plane. Henon's trick multiplies a given dynamical system, $\frac{d\vec{x}}{dt}$, by the reciprocal of the corresponding dynamics of the variable that defines the plane (e.g. $\frac{d\vec{x}}{dt}\frac{dt}{dy}= \frac{d\vec{x}}{dy}$ for PS1). The point exactly on the plane can be identified by first locating the point immediately preceding the plane and then integrating the distance from that point to the control plane, a distance $dx$ for PS0 or a distance $dy$ for PS1. Figure \ref{fig:control_planes} shows each control plane (outlined as a black rectangle) and the corresponding exact points on each plane (yellow for PS0 on the left and purple for PS1 on the right).

\begin{figure}
  \includegraphics[width=\textwidth,height=\textheight,keepaspectratio]{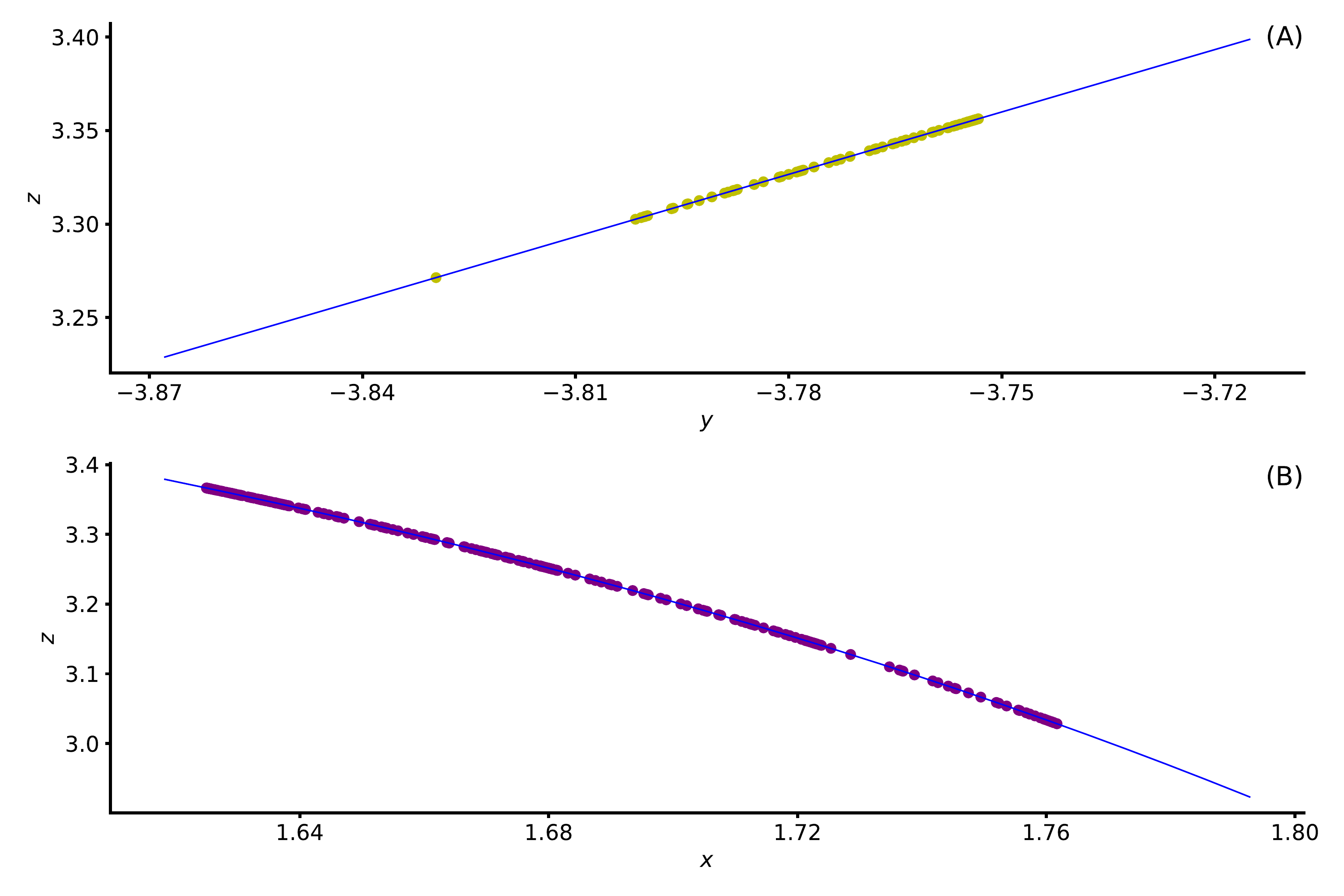}
  \caption{\label{fig:poly_fit}Polynomial fit of the points on each Poincar\'e section. (A) Blue line is the degree 2 polynomial fit, $p_0(\sigma=y)$, of the yellow points (residual squared is $1.71\times 10^{-33}$). (B) Blue line is the degree 3 polynomial fit, $p_1(\sigma=x)$, of the purple points (residual squared is $7.16\times 10^{-15}$).}
\end{figure}

A line of best fit for the points on each control plane is calculated to give a function that approximates the points on the plane. The function $p_i(\sigma)$ is found, for $i=0,1$, that allows for reconstruction of the $z$ value on the control plane from the independent variable. This would give $p_0(\sigma=y)$ for PS0 and $p_1(\sigma=x)$ for PS1. The polynomial allows each control plane to parameterized by 1-dimension.  In Figure \ref{fig:poly_fit}A the resultant line of best fit (degree 2 polynomial) is given where the yellow points are directly on the control plane PS0. The low residual squared ($1.71\times 10^{-33}$) indicates that $p_0(\sigma=y)$ is an excellent approximation to return a $z$ value for a given $\sigma=y$ value on the plane.  In Figure \ref{fig:poly_fit}B the resultant line of best fit (degree 3 polynomial) is given where the purple points are directly on the control plane PS1. The low residual squared ($7.16\times 10^{-15}$) indicates that $p_1(\sigma=x)$ is a very good approximation to return a $z$ value for a given $\sigma=x$ value on the plane.

\begin{figure}
  \includegraphics[width=\textwidth,height=\textheight,keepaspectratio]{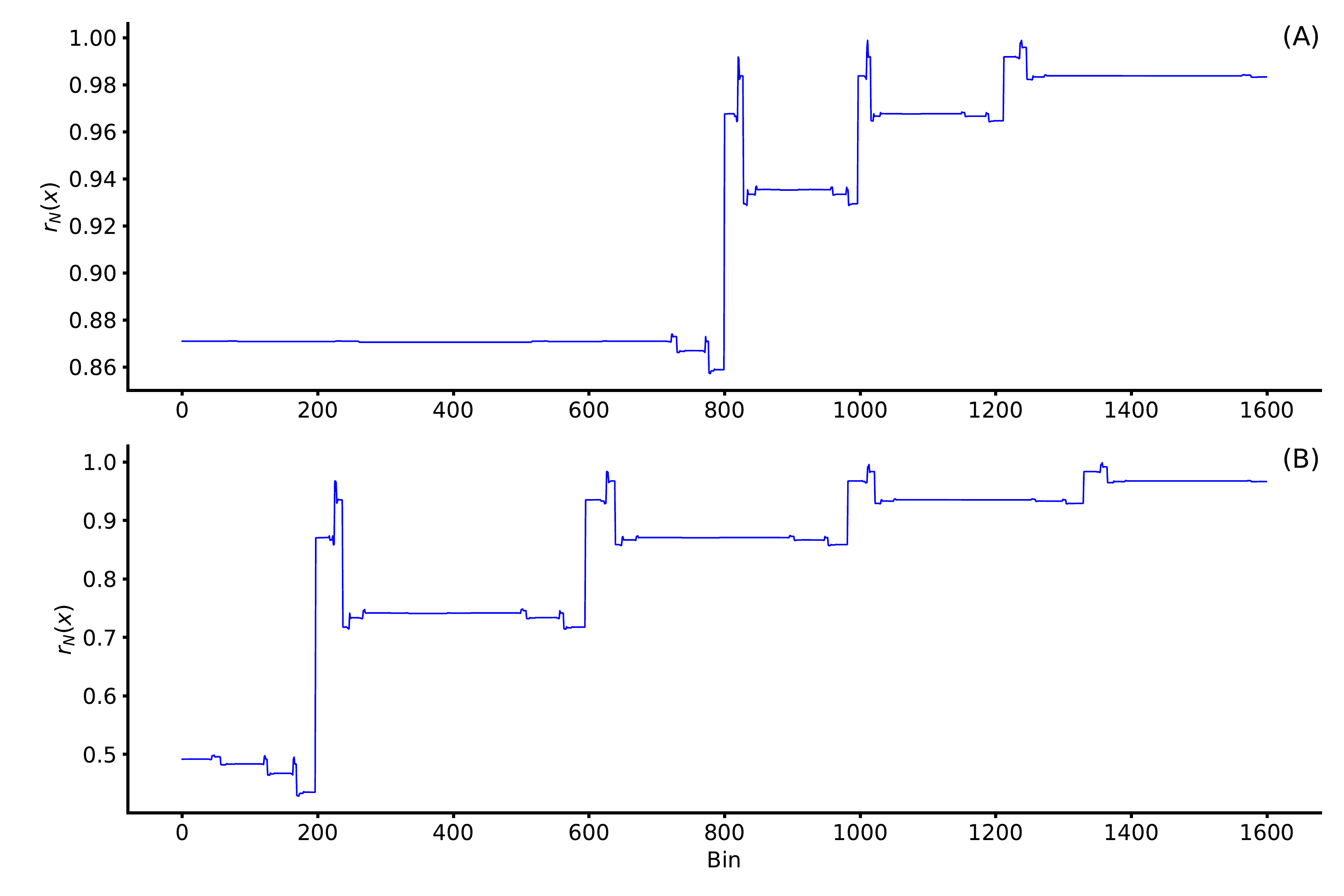}
  \caption{\label{fig:coding_fcn}Coding function $r_N(X)$ with $M = 1600$ bins and $N = 16$ future intersections with the control planes. (A) PS0 coding function where the input $X$ represents the bin center along $p_0(y)$. (B) PS1 coding function where the input $X$ represents the bin center along  $p_1(x)$.}
\end{figure}

After establishing the $p_i(\sigma)$ function for each control plane, the range of values for the independent variable is partitioned into $M=1600$ equally spaced bins. The $y$ values of PS0 are partitioned into $M$ bins and the $x$ values of PS1 are also partitioned into $M$ bins. The middle of each bin is then used as an initial condition of a trajectory and the dynamics are integrated forward in time for each initial condition. A record is made of the binary sequence of control planes that a trajectory visits, also known as the symbolic dynamics, where a 0 indicates the trajectory visited PS0 and a 1 indicates the trajectory visited PS1. The symbolic dynamics are generated for $N$ future iterations around the attractor. As an example, if $N=4$ the symbolic sequence of a bin $B$ on PS0 might be 1011. This sequence means the numerical integration from $B$ results in a trajectory crossing PS1 (1), then PS0 (0), then PS1 twice (11). Each $N$-bit long binary sequence is mapped to a binary decimal and is given as $r_N(X) = \sum_{n=1}^N b_n 2^{-n}$, where $b_1b_2b_3...$ represents the binary symbolic dynamics \cite{Hayes1993}. Figure \ref{fig:coding_fcn} provides a visualization of this coding function with $M=1600$ bins and $N=16$ future intersections with the control planes. For ease of visualization the vertical axis has been converted from the binary decimal to base-10. Each plateau represents identical symbolic dynamics, meaning that for 16 loops around the attractor a trajectory starting in a bin on the plateau visits the same sequence of control planes. Figure \ref{fig:coding_fcn}A plots the PS0 coding function $r_N(X)$ on the vertical axis with the bin number on the horizontal axis. Figure \ref{fig:coding_fcn}B plots the PS1 coding function $r_N(X)$ on the vertical axis with the bin number on the horizontal axis. Figure \ref{fig:coding_fcn} is distinctly different from the coding function of the double scroll oscillator (see \textcite{Hayes1993, Parker1999, Zarringhalam2006, Morena2014a, Morena2014, Morena2020}) that contains many jagged lines and few plateaus.

After establishing the coding function and bin centers for each control plane, a binary control string, $CTRL$, can be applied. This control sequence, $CTRL$, is repeated until a cupolet is formed or the simulation time ends. Each bit of $CTRL$ corresponds to the type of perturbation that is applied when the trajectory intersects with a control plane. As an example, if $CTRL = 1001$ then at the first intersection with a control plane a macrocontrol (1) is applied to the trajectory, then a microcontrol (0) at the next two intersections, followed by a macrocontrol (1) at the fourth intersection. At the fifth intersection, the $CTRL$ sequence resets to the first bit and the process begins again. A microcontrol merely recenters the trajectory to the middle of the bin so the shift is bounded to have a magnitude smaller than $(1/1600)L$, where $L$ is the length of the 1-dimensional function representing the Poincar\'e section. If the bit is a 1 then a macrocontrol is applied. A macrocontrol perturbs the trajectory to a new bin center that has the smallest non-zero difference in symbolic sequence when compared to the symbolic sequence of the current bin center. For $N=16$, this corresponds to a perturbation from bin $x_0$ to a new bin $x_1$ that has a nonzero difference in coding function, $\min| r_N(x_0) - r_N(x_1)| > 0$. Essentially the macrocontrol pushes the trajectory far enough so that it is not on the same plateau (in $r(x)$) as $x_0$.

Due to the bursting nature of the Hindmarsh-Rose dynamics, constraints exist on the possible symbolic dynamics. The uncontrolled symbolic dynamics result in one or more consecutive intersections of PS1 but never more than one consecutive intersection with PS0.

In the next section, this control scheme is applied to Eq \ref{eq:hr} and selected resulting dynamical structures are analyzed.

% CUPOLETS
% Contains Cupolet section of HR Single Cupolet Paper
\section{Cupolets}\label{sec:cupolets}
The key property of the HR neural model we aim to exploit is that that if it receives a sequence of impulses and if those impulses fall into a pattern that is sympathetic with the dynamics of the HR neuron in the chaotic region, then it is possible that the trajectory will be kicked onto a periodic orbit as long as the impulses continue. Furthermore, because of the diversity of trajectories in the chaotic regime, different sets of impulses can lead to different periodic orbits with distinct structure. The model we present here has used idealized impulses, but the impulses serve as proxies for the incoming impulses from nearby neurons.

Several cupolets and cupolet properties are discussed in this section using $M=1600$ bins and $N=16$ for $r_N(X)$. Binary control sequences from 2 bits to 12 bits are used. This results in over 8188 ($\sum_i=2^12 2^i$) total control sequences applied to the Hindmarsh-Rose system. Not every control sequence will result in a cupolet. A cupolet is considered to have formed when the control sequence leads to stabilization of the chaotic trajectory into a persistent, periodic orbit that would otherwise not exist without the control since the system is in a chaotic state.

\begin{figure}
  \includegraphics[width=\textwidth,height=\textheight,keepaspectratio]{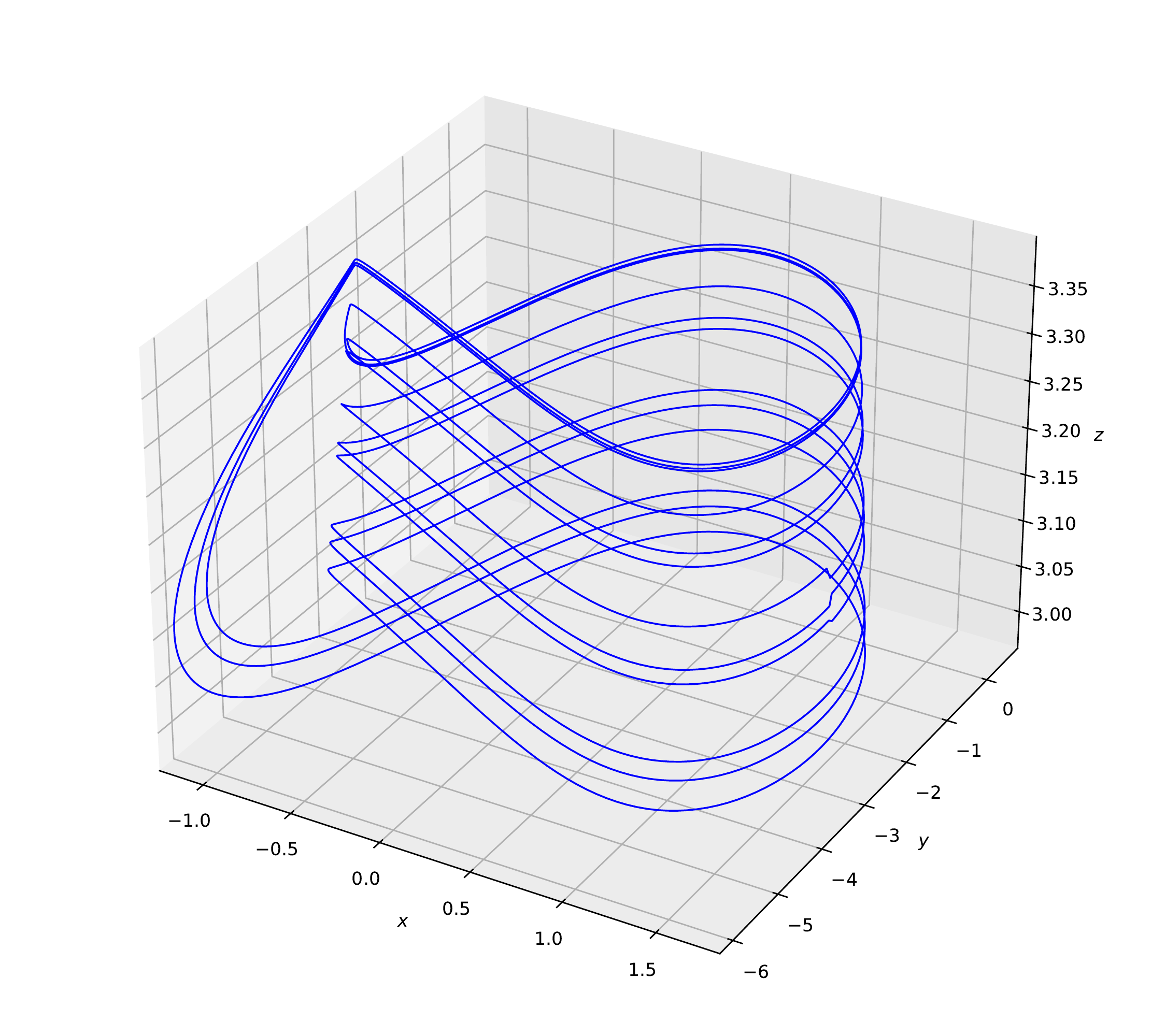}
  \caption{\label{fig:c01011}Cupolet C01011.}
\end{figure}

If a particular control sequence does result in a cupolet, then we have adopted the convention that it is named with a `C' followed by the control sequence. For example, C10 would correspond to the cupolet generated with the control sequence 10. Each cupolet also has a visitation sequence that is the binary string of the symbolic dynamics of the cupolet. The symbolic dynamics reflect the order of control planes that the cupolet visits and is designated by placing a `V' prior to the symbolic dynamics. For example, cupolet C01011 has a visitation sequence of V110111101111011. Due to the implemented control sequence 01011, repeated cyclically, the chaotic Hindmarsh-Rose has collapsed into a persistent, periodic sequence of three four-spike bursts as indicated by the three sets of 1111 (note that when cyclically repeated, the leading and trailing 11 connect). Another example is given in Figure \ref{fig:c01011}, cupolet C01011.

\begin{figure}
  \includegraphics[width=\textwidth,height=\textheight,keepaspectratio]{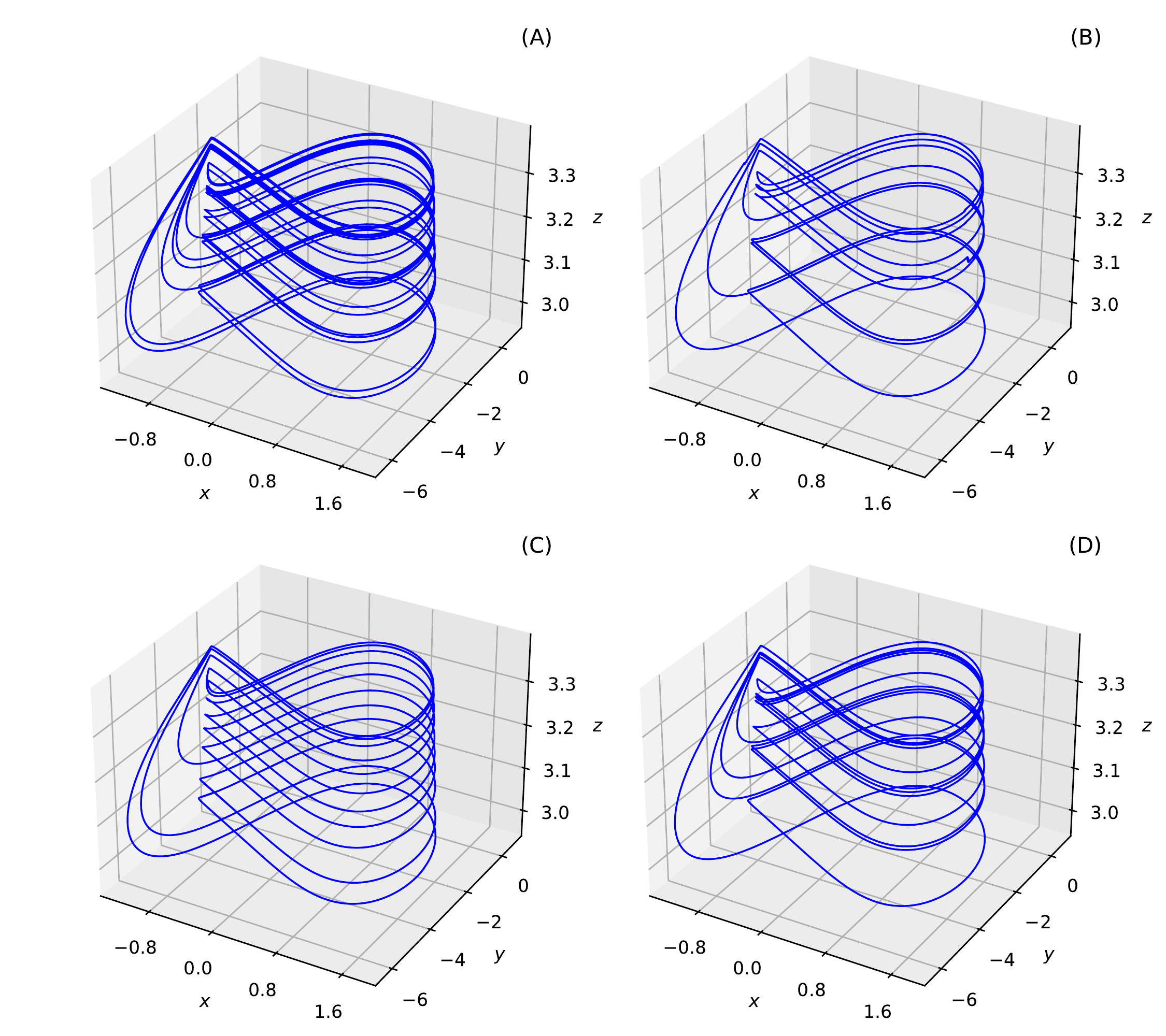}
  \caption{\label{fig:cupolets} (A) Cupolet C11. (B) Cupolet C0110. (C) Cupolet C1010010. (D) Cupolet C01010010.}
\end{figure}

\begin{figure}
  \includegraphics[width=\textwidth,height=\textheight,keepaspectratio]{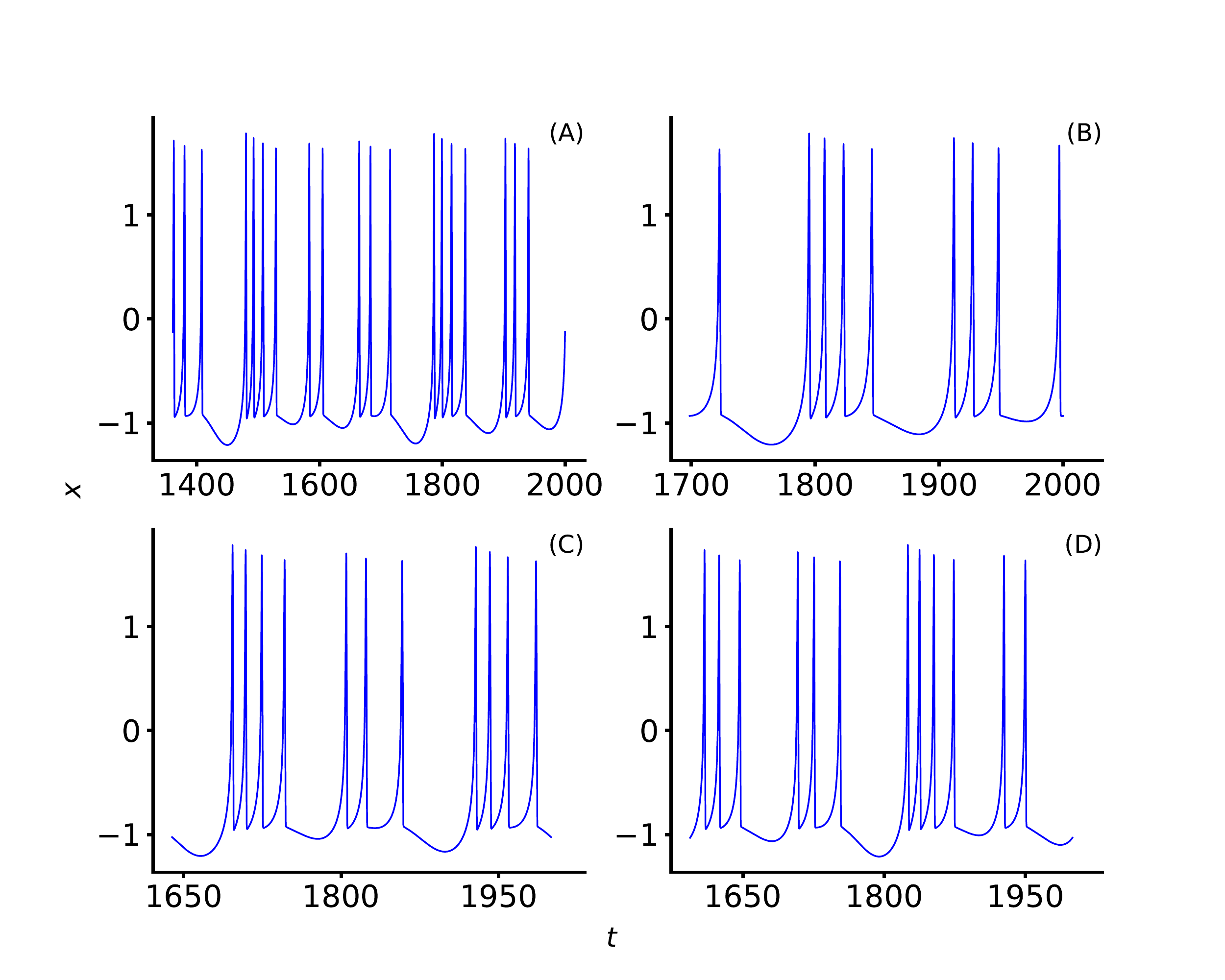}
  \caption{\label{fig:cupolet_analysis} (A) Cupolet C11 single period $x$ time series. (B) Cupolet C0110 single period $x$ time series. (C) Cupolet C1010010 single period $x$ time series. (D) Cupolet C01010010 single period $x$ time series.}
\end{figure}

More cupolet examples are given in Figure \ref{fig:cupolets}. These cupolets are generated from different control sequences. Figure \ref{fig:cupolets}A is C11, Figure \ref{fig:cupolets}B is C0110, Figure \ref{fig:cupolets}C is C1010010 and Figure \ref{fig:cupolets}D is C01010010. Figure \ref{fig:cupolet_analysis} further examines these cupolets by displaying the time series of the $x$ variable which represents the firing pattern of the neuron. Figure \ref{fig:cupolet_analysis}A shows one period of C11. This cupolet has 6 bursts before repeating, where the number of spikes in the bursts are 2, 3, 4, 3, 3, 4. Figure \ref{fig:cupolet_analysis}B shows one period of C0110. This cupolet has 3 bursts before repeating, where the number of spikes in the bursts are 4, 3, 2. Figures \ref{fig:cupolet_analysis}C shows one period of C1010010. This cupolet has 3 bursts before repeating, where the number of spikes in the bursts are 3, 4, 4. Figure \ref{fig:cupolet_analysis}D shows one period of C01010010. This cupolet has 4 bursts before repeating, where the number of spikes in the bursts are 2, 3, 3, 4. Even though these cupolets are created through control of the same chaotic HR neuron, the bursting pattern is not the same. Table \ref{table:cupolets} provides an overview of the patterns seen in each cupolet in Figure \ref{fig:cupolets}. The column for spikes per period, labeled ``Spikes$/T$" is the average number of spikes that occur over one period. For instance, C11 contains 19 spikes in one period while C0110 has 9 spikes. C1010010 has 11 spikes while  C01010010 has 12 spikes in one period. The approximate period of each cupolet has been rounded to two decimal places. The approximate spikes per unit time, or firing rate, is reported in the last column of this table rounded to two decimal places. This collection of cupolets has a wide range of differences in both spikes and period, yet the firing rate remains approximately the same.

\begin{table}
\centering
 \begin{tabular}{||c c c c c c c||}
 \hline
 Cupolet & 2 Bursts & 3 Bursts & 4 Bursts & Spikes & $T$,Period Length & Spikes/$T$\\ [0.5ex]
 \hline\hline
 C11 & 1 & 3 & 2 & 19 & 639.43 & 0.03 \\
 \hline
 C0110 & 1 & 1 & 1 & 9 & 301.25 & 0.03 \\
 \hline
 C1010010 & 0 & 1 & 2 &11 & 360.96 & 0.03 \\
 \hline
 C01010010 & 1 & 2 & 1 & 12 & 406.15 & 0.03 \\[1ex]
 \hline
\end{tabular}
\caption{\label{table:cupolets}Summarized characteristics of a single period of the cupolets in Figure \ref{fig:cupolets}. The ``bursts" column headings indicate the number of spikes in a burst event that occurs in a single period for the cupolet on the left. The period length, $T$, is measured in integration time units and has been rounded to two decimal places. The spikes/$T$, or firing rate, is ratio of spikes to the period length and has been rounded to two decimal places.}
\end{table}

\begin{figure}
  \includegraphics[width=\textwidth,height=\textheight,keepaspectratio]{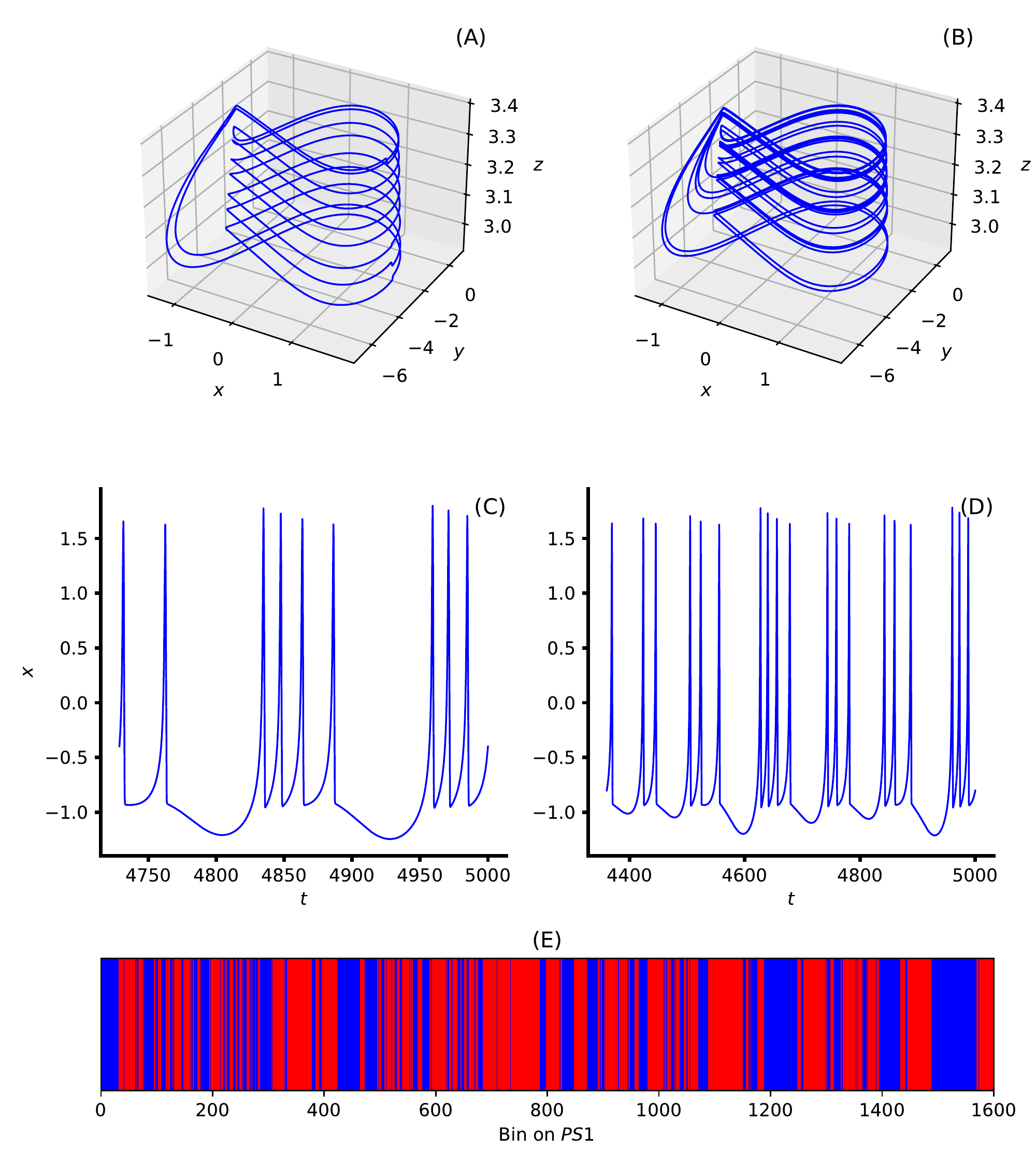}
  \caption{\label{fig:homolog} Homologous cupolets of C11. (A) First type of C11 cupolet, designated C11A. (B) Second type of C11 cupolet, designated C11B. (C) Type C11A single period $x$ time series. (D) Type C11B single period $x$ time series. (E) Graphical depiction of the distribution of initial bins that stabilize onto C11A and C11B. Blue (Red) bar indicates that starting the control sequence in the corresponding bin on the horizontal axis resulted in C11A (C11B). }
\end{figure}

Certain control sequences generate a multiplicity of cupolets, which we will call homologous cupolets. That is, for a given control sequence it is possible that more than one cupolet can be generated. For a given control sequence with homologous cupolets, the specific cupolet generated depends on the initial bin and control plane where the first control in the sequence is applied. For example, C11 has two homologous cupolets, meaning that two unique cupolets result from applying the control sequence 11 repeatedly. The homologous cupolets for C11 are given in Figure \ref{fig:homolog}. Figure~\ref{fig:homolog}A shows cupolet C11A and Figure~\ref{fig:homolog}B shows cupolet C11B. C11A and C11B have distinctly different orbits and spiking behavior. This can be further seen by examining Figure~\ref{fig:homolog}C-D, where the respective $x$ time series for each cupolet is shown.

As can be seen, C11A has fewer spikes (10 spikes) than C11B (19 spikes) in a single period. C11A has a shorter period (271.42 time units) compared to C11B (639.43 time units). The approximate spikes/period are 0.04 for C11A and 0.03 for C11B.

Figure \ref{fig:homolog}E illustrates which bins from PS1 will result in C11A (blue) and C11B (red). Starting in each bin of PS1, the control sequence 11 is applied until a cupolet forms. Then the type of cupolet is checked and if the cupolet is C11A then a blue bar of height 1 is plotted above the corresponding bin number on the horizontal axis. If the resulting cupolet is C11B then a red bar of height 1 is plotted above the corresponding bin number on the horizontal axis. C11A resulted from 42.625\% of the bins, or 682 bins while C11B resulted from 57.375\% of the bins, or 918 bins. This sort of structure and dependence on the fine details of the starting position is reminiscent of fractal basin boundaries in chaos research.

\begin{figure}
  \centering
  \includegraphics[width=\textwidth,height=0.8\textheight,keepaspectratio]{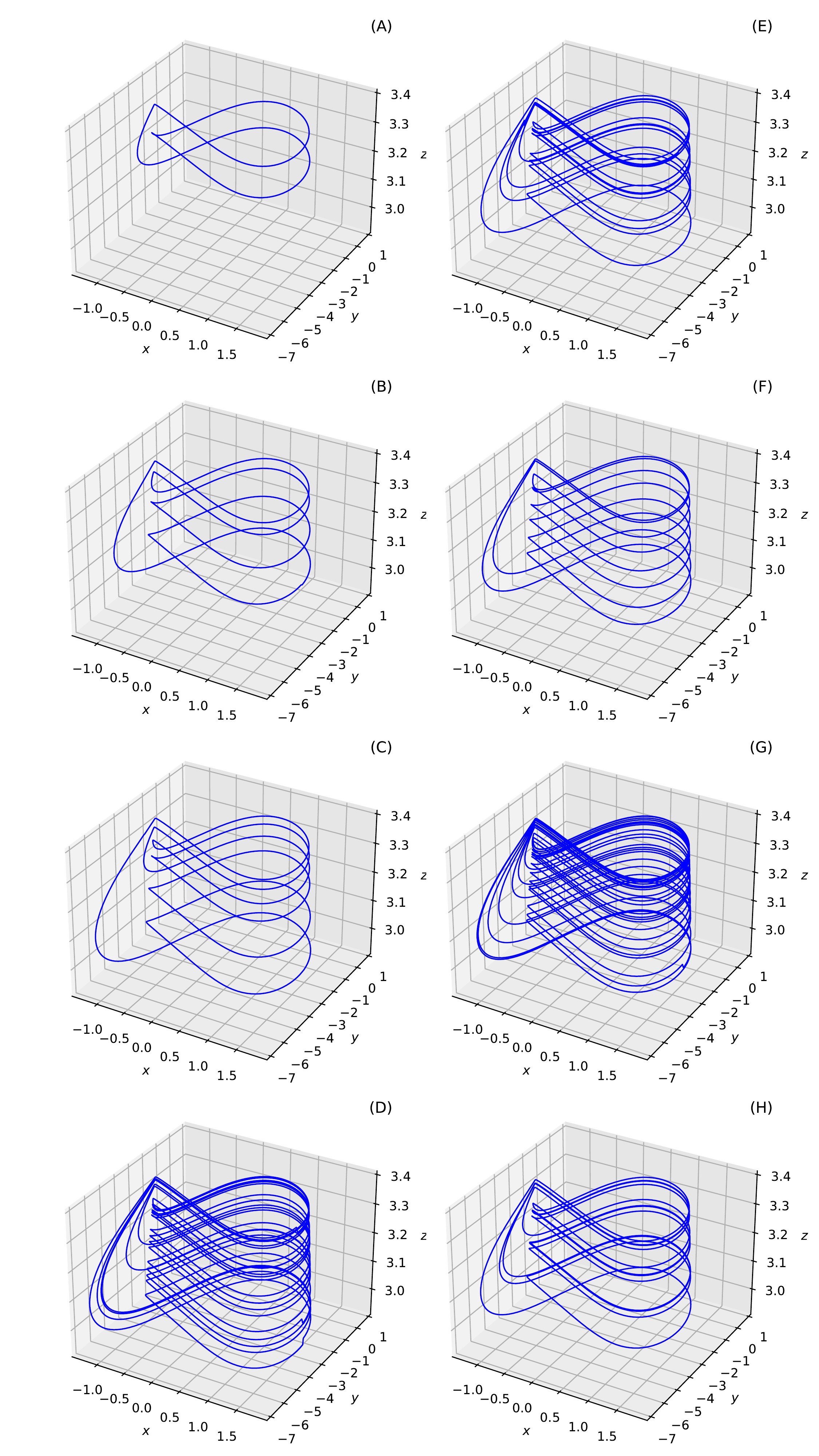}
  \caption{\label{fig:extra_cupolets} Various cupolets resulting from the control scheme. (A) C001, (B) C10010, (C) C11010011, (D) C1101110, (E) C11100010, (F) C10000, (G) C0111110, (H) C01100011.}
\end{figure}

In total 8127 cupolets have been found so far with the control sequence lengths from 2 bits up to 12 bits. Several more examples of cupolets are given in Figure \ref{fig:extra_cupolets} and the properties of these cupolets are given in Table \ref{table:extra_cupolets}.

\begin{table}
\centering
 \begin{tabular}{||c c c c||}
 \hline
 Cupolet & Spikes & $T$,Period Length & Spikes/$T$\\ [0.5ex]
 \hline\hline
 C001 & 2 & 76.35 & 0.03 \\
 \hline
 C10010  & 4 & 137.32 & 0.03 \\
 \hline
 C11010011 & 6 & 196.67 & 0.03 \\
 \hline
 C1101110  & 25 &  697.27 & 0.04 \\
 \hline
 C11100010  & 12 & 408.53 & 0.03 \\
 \hline
 C10000  & 8 & 257.52 & 0.03 \\
 \hline
 C0111110  &22 & 739.86 & 0.03 \\
 \hline
 C01100011 & 12 & 404.05 & 0.03 \\[1ex]
 \hline
\end{tabular}
\caption{\label{table:extra_cupolets}Summarized characteristics of a single period of the cupolets seen in Figure \ref{fig:extra_cupolets}. The spikes are the total number of spikes per period. The period length, $T$, is measured in integration time units and has been rounded to two decimal places. The spikes/$T$, or firing rate, is ratio of spikes to the period length, $T$ and has been rounded to two decimal places.}
\end{table}

% DISCUSSION
% Contains Discussion section of HR Single Cupolet Paper
\section{Discussion}\label{sec:hr_discussion}
\textcite{Rabinovich1999} suggest a role for chaos in the brain and how certain individual neurons in a CPG would exhibit chaotic dynamics if the neuron was not connected. We consider the implications and possible dynamical evolution that might occur if some neurons are in a mathematical parameter state that would exhibit chaotic behavior when no inputs are being received. However, as shown here, the many inputs within the nervous system may stabilize the chaos into a regular pattern. The different binary control sequences act as an analog for the myriad biological inputs a neuron might receive. In \S\ref{sec:cupolets}, we show that many binary control sequences can cause the neuron to stabilize onto a cupolet. However, we also found that not all control sequences will result in a cupolet and some control sequences result in homologous cupolets. Therefore, a neuron may have many unique behaviors, corresponding to unique cupolets, yet the system parameters remain unchanged and the HR neuron would be chaotic without the stabilization induced by the controls. The controls are convenient ``stand-ins" or representations of the possible incoming spike-like signals that a neuron may receive.

\textcite{Morena2020} provide an analysis of the cupolets from the double scroll system and show that certain cupolets are in some sense fundamental. These fundamental cupolets can then be used to generate more complicated composite cupolets. Isolating the fundamental cupolets within the chaotic Hindmarsh-Rose system may provide further insight into the potential dynamics of the system and properties of neural dynamics.

In this paper we reported on the existence of cupolets within a chaotic neural system, the Hindmarsh-Rose dynamical model. Even though there are no direct biological parameters in the model, the dimensionless $x$ variable represents observed neural firing patterns, specifically bursting behavior. \textcite{Izhikevich2004} discusses the usefulness of certain neural models and reports that the Hindmarsh-Rose model has the capability to generate many observed neural firing patterns. This suggests that cupolets may exist biologically, where the controls are applied through auxiliary cells or input impulses from connected neurons. \textcite{Morena2014} demonstrated how interacting cupolets in the double scroll oscillator can mutually stabilize through an interaction function. For certain interaction functions, and certain cupolets, this communication results in the two cupolets locking into a periodic, persistent interaction with no external controls, in a state of chaotic entanglement. One interaction function used in \textcite{Morena2014} was based on integrate-and-fire neural dynamics. This interaction function motivated the result in \textcite{Parker2020}, where it was shown that two bidirectional neurons, each modeled as an adapted FitzHugh-Nagumo model \cite{FitzHugh1961, Nagumo1962}, were capable of transitioning to a mutually stabilized periodic state from a chaotic state in the presence of a certain external signal, although the system involved synaptic learning. This suggests a future research investigation examining whether a bidirectional Hindmarsh-Rose model may be able to achieve the mutual stabilization shown in \textcite{Morena2014} and \textcite{Parker2020}. The advantage of the Hindmarsh-Rose model over the FitzHugh-Nagumo model is that each individual neuron can exhibit chaotic behavior leading to a more direct comparison with the mutual stabilization shown in \textcite{Morena2014}.

We note this is an initial report of cupolets in a chaotic Hindmarsh-Rose system. It is reasonable to hypothesize that cupolets exist in a higher dimensional neural model capable of chaos, for example the Hodgkin-Huxley model. A control scheme similar to that which has been presented here has not been attempted on a chaotic system with 4-dimensions, so it will be an interesting topic for future research to see if generation of cupolets is possible.

% REFERECNES
\printbibliography

\end{document}